**Probing the Electronic Properties of Monolayer MoS$_2$ via Interaction with Molecular Hydrogen**

*Natália P. Rezende, Alisson R. Cadore, Andreij C. Gadelha, Cíntia L. Pereira, Vinicius Ornelas, Kenji Watanabe, Takashi Taniguchi, André S. Ferlauto, Ângelo Malachias, Leonardo C. Campos, and Rodrigo G. Lacerda\**

N. P. Rezende, Dr. A. R. Cadore, A. C. Gadelha, C. L. Pereira, V. Ornelas, Prof. A. Malachias, Prof. L. C. Campos, Prof. R. G. Lacerda*
Departamento de Física, Universidade Federal de Minas Gerais, Belo Horizonte, 30123-970, Brasil
E-mail: rlacerda@fisica.ufmg.br
Prof. A. S. Ferlauto
Centro de Engenharia, Modelagem e Ciências Sociais Aplicadas, Universidade Federal do ABC, 09210580, Brasil
 K. Watanabe, T. Taniguchi,
Advanced Materials Laboratory, National Institute for Materials Science, Namiki, 305-0044, Japan

Keywords: hydrogen detection, monolayer MoS$_2$, gas interaction, field effect transistors

This work presents a detailed experimental investigation of the interaction between molecular hydrogen (H$_2$) and monolayer MoS$_2$ field effect transistors (MoS$_2$ FET), aiming for sensing application. The MoS$_2$ FET exhibit a response to H$_2$ that covers a broad range of concentration (0.1-90 %) at a relatively low operating temperature range (300-473 K). Most important, H$_2$ sensors based on MoS$_2$ FETs show desirable properties such as full reversibility and absence of catalytic metal dopants (Pt or Pd). The experimental results indicate that the conductivity of MoS$_2$ monotonically increases as a function of the H$_2$ concentration due to a reversible charge transferring process. It is proposed that such process involves dissociative H$_2$ adsorption driven by interaction with sulfur vacancies in the MoS$_2$ surface ($V_S$). This description is in agreement with related density functional theory studies about H$_2$ adsorption on MoS$_2$. Finally, measurements on partially defect-passivated MoS$_2$ FETs using atomic layer deposited aluminum oxide consist of an experimental indication that the $V_S$ plays an important role in the H$_2$ interaction with the MoS$_2$. These findings provide insights for futures applications in catalytic process between monolayer MoS$_2$ and H$_2$ and also introduce MoS$_2$ FETs as promising H$_2$ sensors.

1.  Introduction

Molecular hydrogen (H$_2$) has great potential as a non-polluting source of energy used in several industrial applications. Especially it is used for desulphurization of petroleum products, synthesis of ammonia, methanol and as a green fuel in mobile applications such as cars and rockets.[1] For these reasons, areas involving H$_2$ generation, storage and detection are under continuous development. In special, sensing of H$_2$ is crucial for safety applications due to its flammability range of 4.0-75.0 % by volume in the air.[1,2] Hence, it is indispensable to develop high-performance sensors with a wider measurement range (1-99% v/v $H_2$) to detect and monitor $H_2$ for industrial and safety applications.[2] There are several commercially available hydrogen sensors classified by the sensing mechanism like electrochemical, catalytic, work-function-based and resistance-based.[1] However, most of these sensors present limitations that need to be overcome to satisfy the requirements for specific applications. For example, gas chromatography and mass spectroscopy are usual systems for $H_2$ detection that have an inherent drawback of large hardware size, requiring expensive and constant maintenance.[1,2] Resistive $H_2$ sensors based on metal oxide semiconductors generally operate at high temperatures (between 453 K and 723 K) and need a fraction of oxygen in the ambient to detect $H_2$.[1,3] For this reason, there is an increasing demand for the development of $H_2$ sensors that are compact, operate at room temperature, with low cost and improved performance.

Recently, two-dimensional (2D) layered materials have attracted scientific interest due to their properties such as high surface area, low electrical noise, and high electrical conductivity.[4] With considerable impact on gas sensor applications, 2D materials have their electric conductivity depending on the chemical state of their surface, which drastically changes under adsorption of gas molecules.[4–6] Such property, for instance, is exploited in graphene researches for detection of distinct gases such as $NH_3$,[7] $NO_2$,[8] $CO_2$[9] and others. In this context, molybdenum disulfide ($MoS_2$) is one of the most promising transition metal dichalcogenides with great semiconducting properties, including a large intrinsic band gap of 1.8 eV for the monolayer and high current on/off ratios.[10,11] Few-layer $MoS_2$ field effect transistors have been particularly successful for applications in nanoelectronics, optoelectronics and gas detection.[12,13] In fact, $MoS_2$ transistors have been used to monitor gases such as $O_2$,[14] NO,[15] $NH_3$,[16,17] $NO_2$,[16–18] and, in general, sensors based on few layers $MoS_2$ FETs exhibit advantages comparing with bulk sensors such as transparency, cost-effectiveness for massive production and high sensitivity.[15,16,19] In terms of $H_2$ sensors by non-functionalized $MoS_2$ monolayer FETs ($MoS_2$ FET), the mechanism of detection has not been completely understood yet. Some works propose the $H_2$ detection mechanism that holds only for more complex structures like $MoS_2$ nanocomposites films doped with palladium (Pd),[20] platinum (Pt)[21] or heterojunctions of $MoS_2$ films and silicon.[22] Recently, Agrawal *et al*.[23] have shown a promising $H_2$ detection system using edge-oriented vertically aligned $MoS_2$ flakes in a three-dimensional array without any doping, suggesting that bare $MoS_2$ has the potential for $H_2$ detection.

Here, we present a systematic study of the electronic properties of $MoS_2$ FETs under $H_2$ gas exposure. The $MoS_2$ sensors can detect a wide range of $H_2$ concentration (0.1 % to up to 90 %) as well as operate even at room temperature (300 K). These two characteristics are important for $H_2$ sensors in practical applications. Our results indicate that the $H_2$ reaction is independent of the choice of the substrate or metallic contacts. Based on our experimental data and previous density functional theory (DFT) studies, we suggest a model able to explain the electronic response of $MoS_2$ under interaction with $H_2$. We believe that $H_2$ molecules dissociate on the S vacancies ($V_S$) of $MoS_2$ transferring charge to $MoS_2$.[24–26] This report provides strong evidence that $MoS_2$ helps the $H_2$ catalytic reaction with clear implications for hydrogen molecular sensors. Additionally, the understanding of $H_2$ interaction with $MoS_2$ is also strategic for industrial processes, such as hydrogen storage and hydrodesulphurization.

## 2. Results and Discussion

After thermal annealing for 12 hours in ultra-high pure Ar, we investigate the response of the $MoS_2$ FET supported on $SiO_2$/Si substrate for molecular hydrogen concentration (*[H₂]*) of 20 % at 473 K. The optical image of the $MoS_2$ FET is shown in **Figure 1**a. We present the results at such temperature (473K) because of the $MoS_2$ FET exhibit larger response to the $H_2$, where faster desorption of $H_2$ is also obtained, as discussed in detail along this work. Initially, we carried out source-drain current ($I_{SD}$) vs.

gate voltage ($V_G$) measurements at fixed source-drain voltage $V_{SD} = 1$ V for distinct $H_2$ exposure times, as shown in Figure 1b. The cyan (lower) $I_{SD}$ vs. $V_G$ curve represents the standard current ($I_{SD}^{INIT}$) without the presence of $H_2$ and the blue (upper) $I_{SD}$ vs. $V_G$ curve corresponds to the saturation after 45 min of $H_2$ exposure ($I_{SD}^{H_2}$). The saturation criterion used here implies that the transfer curves became constant under $H_2$ exposure. The characteristics of the $I_{SD}$ vs. $V_G$ curves in Figure 1b represents the typical n-type nature of the $MoS_2$ channel, which the current increase with the gradual increase in $V_G$ due to the accumulation layer of electrons.[11] This observation is consistent with previous works, which report that the $MoS_2$ transistor is predominantly n-type duo to the spontaneous doping promoted by the sulfur vacancies.[27] Furthermore, note that the inset of Figure 1b shows a linear (ohmic) dependence of the $I_{SD}$ vs. $V_{SD}$ without $H_2$ exposure, excluding the possibility of the Schottky barrier as the dominant transport mechanism.[11]

In all devices that we measured, the presence of molecular hydrogen causes a leftward shift in the $I_{SD}$ vs. $V_G$ curves, with a simultaneous increase of the $I_{SD}$ current. This modulation of the $MoS_2$ conductivity indicates a strong response of the $MoS_2$ FET to $H_2$. This leftward shift means that is necessary a more negative $V_G$ turn off the FET, which indicates an increase in the density of $MoS_2$ carriers after $H_2$ exposure. Besides, the conductivity change ($\Delta\sigma$) of the $MoS_2$ channel is expressed by the equation: $\Delta\sigma = e\mu\Delta n$, where e is the electron charge, n is the charge carrier density, and µ is the charge mobility. Thus, a conductivity increase can happen with an increase of charge carrier density or mobility or a combination of both. We discuss what is happening with these properties individually in more detail below.

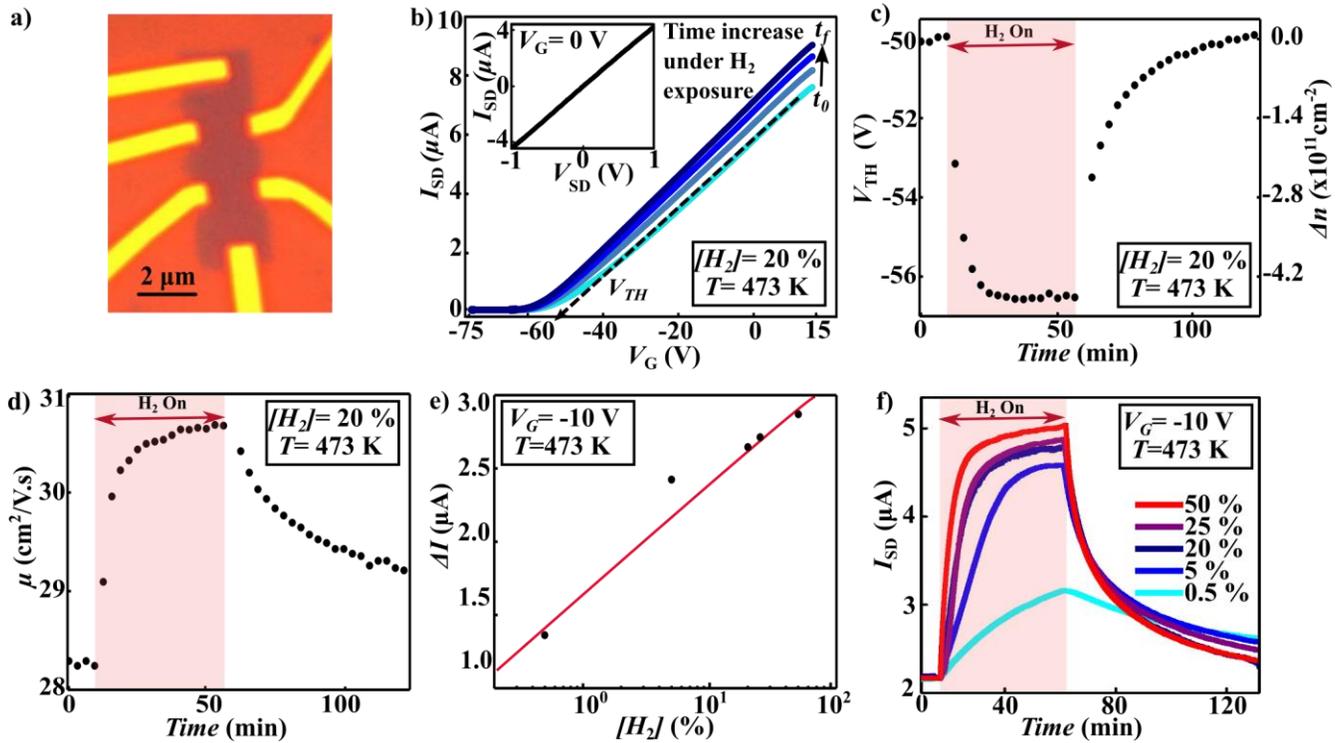

**Figure 1.** Study of $H_2$ interaction with $MoS_2$ FET at 473 K and $V_{SD}$ = 1 V. a) Optical image of a typical $MoS_2$ FET. In b), c) and d) the data were taken at a fixed hydrogen exposure of $[H_2]$= 20 %. b) $I_{SD}$ vs. $V_G$ curves of the $MoS_2$ FET for distinct $H_2$ exposure times. Inset: $I_{SD}$ vs. $V_{SD}$ curve of the $MoS_2$ FET for $V_G$= 0 V, before $H_2$ exposure. c) The left y-axis shows the threshold voltage ($V_{TH}$), and the right y-axis axis shows the amount of charge transferred ($\Delta n$) to $MoS_2$ FET as a function of time under $H_2$ exposure ($H_2$ ON) and its release in pure Ar. d) FET mobility ($\mu$) as a function of time under $H_2$ exposure ($H_2$ ON) and $H_2$ release. e) Change of drain current $\Delta I = I_{SD}^{H_2} - I_{SD}^{INIT}$ as a function of different $H_2$ concentrations at $V_G$= -10 V.  f) Drain current ($I_{SD}$) versus time for different $H_2$ concentrations at $V_G$ = -10 V.

In Figure 1c we plot the values of the threshold voltage ($V_{TH}$) in the left y-axis as a function of $H_2$ exposure time. The $V_{TH}$ is obtained by extrapolation of the linear region of the $I_{SD}$ vs. $V_G$ curves presented in Figure 1b. Once the $H_2$ is introduced into the test chamber, $V_{TH}$ shifts towards negative voltages, indicating that a charge transfer process is taking place due the $H_2$ adsorption. In this case, a fraction of the $H_2$ molecules donate electrons to the $MoS_2$ channel. Next, we shut off the $H_2$ gas (desorption process), and the threshold voltage returns to its initial value, recovering its condition without $H_2$ exposure ($V_{TH}^{INIT}$), indicating the reversibility of the process. The charge transfer process observed is different from what was reported by Agrawal *et al*.[23] where the $H_2$ molecules receive electrons from MoS2. However, our observation is consistent with other $MoS_2$ film sensors doped with Pd, Pt, and heterojunctions of $MoS_2$ films and silicon, which propose that hydrogen acts as a reducing agent donating electrons to the $MoS_2$. [18,20–22]

The amount of charge per unit of area transferred from hydrogen to the $MoS_2$ channel *(Δn)* is shown in the right y-axis of Figure 1c. $\Delta n$ is evaluated by the following equation $\Delta n = c/e \, (V_{TH}^{H_2} - V_{TH}^{INIT})$, where $c = \varepsilon\varepsilon_0/ed$ is the gate capacitance per unit area (12 nF/cm$^2$ for 285 nm of $SiO_2$), $e$ is the electron charge and $V_{TH}^{H_2}$ is the threshold voltage under the $H_2$ exposure. Interestingly, the negative charge transfer from the $H_2$ molecules saturates in approximately 4.5x10$^{11}$ cm$^{-2}$, suggesting that there is a limited number of active sites of the $MoS_2$ surface for the $H_2$ interaction. In addition, we show that the FET mobility ($\mu$) changes during $H_2$ exposure and desorption in Figure 1d. Noticeably, there is an enhancement of the FET mobility during the adsorption of hydrogen. We believe that such behavior may occurs due to neutralization of charged impurities of the $MoS_2$ device under $H_2$ interaction, suggesting a reduction in the charge scattering mechanism. Thus, the conductivity increase in the $MoS_2$ FET is due to a combination of the increase of charge carrier density and mobility.

Further, we present a study of the $MoS_2$ FET response concerning molecular hydrogen concentration. Figure 1e shows a linear relationship between the device current $\left(\Delta I = I_{SD}^{H_2} - I_{SD}^{INIT}\right)$ and *[$H_2$]* in a semi-log scale, using a fixed $V_{SD}$= 1 V and $V_G$= -10 V for 0.5, 5, 20, 25 and 50 % of *[$H_2$]*. Such property can be used as a method to determine the *[$H_2$]* inside the test chamber. In Figure 1f we show the $I_{SD}$ vs. *Time* curve. Note that there is a considerably increase in $I_{SD}$ that is strongly related to the hydrogen concentration. Additionally, in the Supplementary Information, we present measurements for another $MoS_2$ FET in the range of concentrations spanning from 0.1 % up to 90 % of $H_2$. Hydrogen concentrations below 0.1 % could not be measured due to limitations of our experimental setup. The mass flow controller used in our experimental setup does not produce reliable values for *[H$_2$]* below such amount. However, the device response indicates that the lower bound of the H$_2$ detection limit is actually lower than our reports. These results demonstrate that the MoS$_2$ FET is able to detect a large range of H$_2$ concentrations, as it is required for hydrogen sensors. Usually, the response time ($T_{RES}$) of a sensor is determined by the time required to reach 90 % of the total conductance change under the gas exposure and the recovery time ($T_{REC}$) is the time necessary for the current to recover 90 % of its ground state.[20,22] The minimal $T_{RES}$ is found to be approximately 7 min under 50 % of H$_2$ (see Figure S1 in Supplementary Information), while the $T_{REC}$ is around 67 min for the same concentration. The response time of a sensor decreases for higher concentrations of hydrogen *[H$_2$]*. These high recovery and desorption times indicate that the H$_2$ interaction with the MoS$_2$ FET is limited by a diffusion reaction. This characteristic will be discussed in detail later.

Next, we show our investigation on the effects of temperature on the hydrogen detection with MoS$_2$ FETs. In **Figure 2**a, we show the MoS$_2$ FET sensor response (*S*) under *[H$_2$]*= 20 % exposure from 300 K up to 473 K. The sensor response rises up with temperature indicating that the process is thermally activated. Also, we observe a dependency of the device recovering time with temperature. The rate of desorption is slower at 300 K, compared to the rate of desorption at 473 K, meaning that desorption is also a thermally activated process, see Figure 2b. The inset in Figure 2b shows the recovery percentage of intrinsic properties of the MoS$_2$ FET *(Rec)* after one hour of H$_2$ desorption as a function of the temperature, where the red curve is used only as a guide to the eyes. This data supports the reliability of the annealing of MoS$_2$ FET at 473 K for a couple of hours as an efficient method to speed up the system recovery. Also, thermal annealing is frequently used to restore gas sensors based in 2D materials.[7,14] The sensor response is generally defined as $S = (I_{SD}^{H_2} - I_{SD}^{INIT})/I_{SD}^{INIT}$.[14,15] The sensor response as a function of the temperature is obtained by $I_{SD}$ x $V_G$ curves measured with fixed $V_G$= -10 V. Nevertheless, the values of $S$ retrieved in this configuration show the same behavior for other gate values.

For a better understanding of the reversibility of the process at 473 K, we carried out experiments with longer $H_2$ exposure times (see Figure S2 in Supplementary Information). We observe that for 473 K, the $MoS_2$ FET operates reproducibly, showing the same response repeated times. The initial conditions are fully recovered after a couple of hours without $H_2$ interaction in all cases. Such full recovery indicates that neither permanent bonds nor defect creation occur between $H_2$ molecules and $MoS_2$. Indeed, we confirmed this assumption by performing Raman spectroscopy of the monolayer before and after $H_2$ exposure (Figure S3 in Supplementary Information). We do not observe any indication of structural changes or irreversible chemical bonding after exposing the devices to $H_2$, which should be detected by changes in the Raman spectrum.[28] More precisely, no change of the Raman peaks $E_{2g}^1$ and $A_{1g}$ of the $MoS_2$ is observed.

Subsequently, we investigate the dependence of the sensor response with the applied gate voltage. Previous works showed that the response of sensors based in $MoS_2$ for $O_2$,[14] $NO_2$ and $NH_3$[16] detection is improved in a given range of gate voltages. We also observed this behavior for hydrogen sensor response in our devices, as shown in Figure 2c. $S$ becomes remarkably higher for negative gates. The inset of Figure 2c shows that the $V_{TH}$ for this device is approximate -16 V, when the voltage approaches this value, $S$ increases rapidly. This means that the introduction of $H_2$ in the off-state of the transistor makes the current change more significantly than in the condition with the $MOS_2$ FET initially in on-state

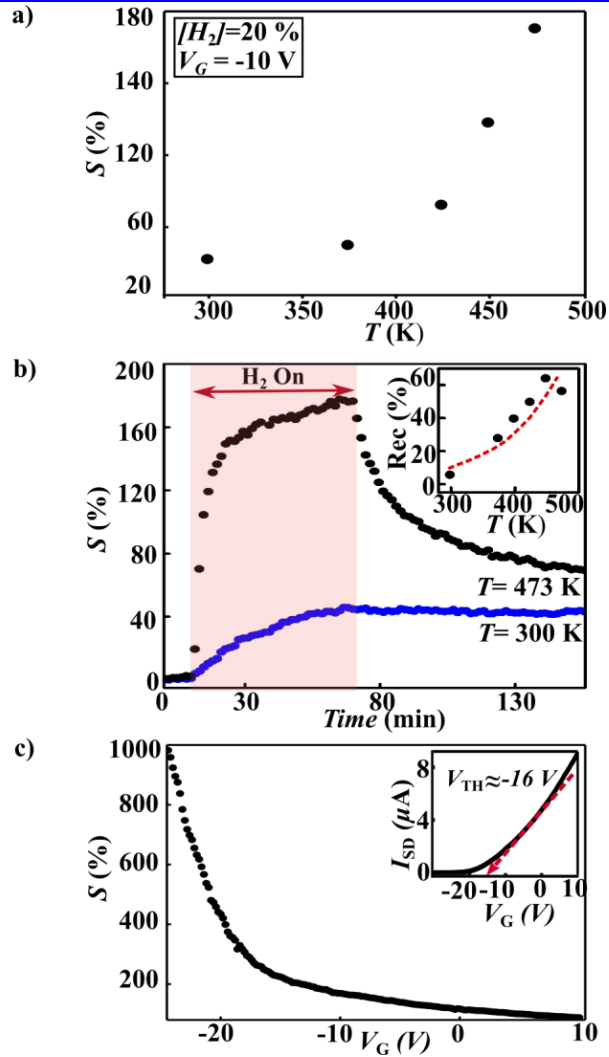

**Figure 2.** Sensing performance of MoS$_2$ FET for a fixed H$_2$ concentration of $[H_2]$= 20 % and $V_{SD}$= 1 V. In a) and b) the data were taken at a fixed $V_G$= -10 V. a) Sensor response ($S$) of MoS$_2$ FET as a function of temperature: from 300 K up to 473 K. b) Comparison of sensor response as function of time under the H$_2$ exposure (H$_2$ ON) at 300 K (black curve) and 473 K (blue curve). Inset:  Recovery ($Rec$) percentage after one hour of H$_2$ desorption as a function of the temperature. c) Sensor response as a function of gate voltage at 473 K. Inset: $I_{SD}$ vs. $V_G$ curve of the MoS$_2$ transistor.

Now, we show our investigation trying to elucidate how MoS$_2$ FETs detect molecular hydrogen. In this context, we carry out experiments specifically designed to clarify: (1) If there is the detection of hydrogen at the MoS$_2$/metal interface.  For example, there are H$_2$ sensors based on the modulation of the electrostatic properties of heterojunctions between the metal contact/2D materials and sensors based on the Schottky barrier effect.[1,16,29–31] (2) If the detection of H$_2$ depends on the underlying substrate.

Previous reports have shown that oxygen gas could interact with graphene devices on $SiO_2$/Si substrates via the incorporation of $O_2$ molecules between the substrate and the graphene causing a change in its FET mobility.[32] (3) And, finally, if the $H_2$ detection is mainly related to intrinsic properties of $MoS_2$. In this case, the mechanism is driven by dissociative adsorption of hydrogen at sulfur vacancies of $MoS_2$. This possibility could explain there is a charge transferring to $MoS_2$ under interaction with $H_2$. Below we discuss each of these points separately.

To check if there is the detection of hydrogen at the interface between $MoS_2$ and the contact metal, or if the contact resistance changes under interaction with $H_2$, we perform measurements in two and four-probe (Hall bar) configurations. It is reasonable to assume that the total change in conductance ($\Delta G$) of $MoS_2$ FET due to $H_2$ adsorption is represented by $\Delta G = \Delta G_{Channel} + \Delta G_{Contacts}$, where $\Delta G_{Channel}$ is directly proportional to the carrier concentration injected in the channel and $\Delta G_{Contacts}$ is inversely proportional to the contact resistance change. Besides, it is known that the four-probe configuration eliminates the influence of contact resistance; in this case, we can detect only the change in conductance due to the charge transfer to the $MoS_2$ channel $(\Delta G = \Delta G_{Channel})$.[14,31] Therefore, if $H_2$ sensing is dominated by the contact resistance, i.e., $H_2$ molecules change the Schottky barrier height via modulation of the metal work function, we would not expect to measure a significant hydrogen response in a four-probe configuration. **Figure 3**a shows the sensor response as a function of the $H_2$ exposure measured in two and four-probe configurations, represented by the black and green curve, respectively (more details are provided in the Supplementary information). Since *S* exhibits the same magnitude for both configurations, we infer that the sensing mechanism does not depend on the electrostatic configuration at the $MoS_2$/metal contact interface. This is also an expected result since our Au/$MoS_2$ contacts have ohmic behavior as depicted in the inset of Figure 1(b).

This is the authors' version (pre peer-review) of the manuscript: N R. Rezende et al, Advanced Electronic Materials  https://doi.org/10.1002/aelm.201800591 , That has been published in its final form:  https://onlinelibrary.wiley.com/doi/abs/10.1002/aelm.201800591Next, we consider the influence of the underline substrate on the $H_2$ detection by $MoS_2$ FETs. Trying to understand the role of the substrate in the hydrogen response, we produced a device with $MoS_2$ supported on h-BN. The h-BN has an inert and flat surface that enables the creation of only a few charge impurities, in contrast to the abundance of trap states that are present in the interface with $SiO_2$/Si substrates.[33,34] We suspected that $H_2$ could interact with substrate defects and dangling bonds causing changes in $MoS_2$ electronic mobility, such as those shown in Figure 1d. In this case, the total change in the $MoS_2$ conductance due to $H_2$ adsorption would be represented by $\Delta G= \Delta G_{Channel} + \Delta G_{SiO2}$, where $\Delta G_{SiO2}$ represents the increase in the conductance due to the interaction or screening of defects of the substrate after the hydrogen passivation. In Figure 3b we show the comparison of the sensor response of the $SiO_2$/Si (black curve) and h-BN substrate (green curve) as a function of the threshold voltage difference ($V_G$-$V_{TH}$). Such normalization (plotted in the x-axis) is adopted because the devices have different $V_{TH}$ and, as previously shown, $S$ depends on the gate voltage. These results reveal that the sensor response is nearly the same for both substrates providing strong evidence that the sensing mechanism does not involve possible interactions between $H_2$ molecules with the underlying substrate. A more complete overview of $MoS_2$ FET supported on h-BN interaction with $H_2$ is found in the Supplementary Information (Figure S4). All fabricated devices present the same sensor response as a function of temperature and $H_2$ concentration as the $MoS_2$ supported on $SiO_2$/Si.

Based on the analysis presented above, we can assert that the $H_2$ sensing mechanism occurs in the $MoS_2$ main channel. Therefore, at this point, we need to elucidate how the interaction of $H_2$ with the $MoS_2$ results in the conductance change of the $MoS_2$ in a reversible process. Some hypothesis can be drawn to explain the conductivity enhancement based in previously reported studies.[35,36] The main mechanism proposed here is the dissociative adsorption of hydrogen facilitated by the catalytic properties of MoS$_2$, as illustrated in Figure 4. The $\Delta G$, in this case, is driven by the charge transfer mechanism from H$_2$ molecules to the MoS$_2$ channel, represented by the following H$_2$ reaction: $H_2 \xrightarrow{MoS_2} 2H^+ + 2e^-$. In this scenario, absorbed H atoms donate electrons to the conduction band of MoS$_2$, producing the sensor response. Theoretical and experimental works report that the existence of $V_S$ increase the catalytic activity of MoS$_2$.[37,38] These references show that the hydrogen evolution is effectively improved due to the creation of $V_S$ in the monolayer MoS$_2$.[37,38]

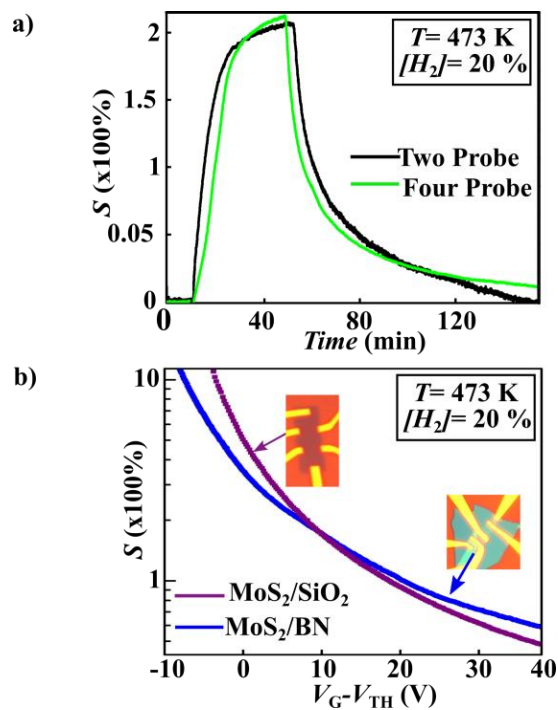

**Figure 3.** Influence of MoS$_2$ FET interfaces on the H$_2$ sensing mechanism. Sensing performance of MoS$_2$ monolayer transistor in a) and b) study for the H$_2$ concentration of *[H$_2$]= 20%* and 473 K. a) Comparison of MoS$_2$ FET sensor response in two-probe and four-probe (Hall bar) measurements. b) Sensor response versus threshold normalized voltage ($V_G$-$V_{TH}$) for MoS$_2$ FETs supported on SiO$_2$/Si and the h-BN substrate.

Now we discuss in details the H$_2$ dissociative adsorption regarding absence and presence of sulfur defective sites in the MoS2 surface. If a perfect MoS$_2$ monolayer is considered, H atoms can bond to the sulfur atoms or molybdenum atoms after H$_2$ dissociation. Theoretical works show that H atoms prefer to bond to sulfur atoms, presenting relatively lower energy concerning Mo atoms. The adsorbed H atoms affect the electronic property of MoS$_2$, promoting a metallization of the surface due to the electron donation.[24,36,39] Yakovkin *et al.*[39] have investigated the relative instability of this bond, and suggested that some H atoms can overcome the potential barrier for desorption and form $H_2$ molecules at high temperature.

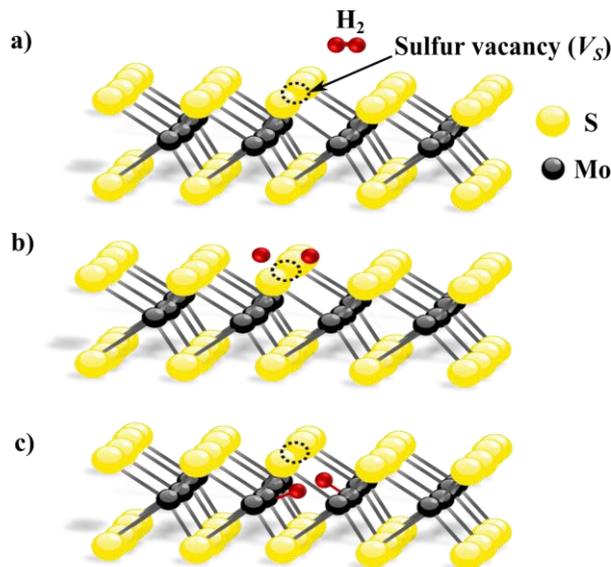

**Figure 4:** Representation of the dissociative adsorption of $H_2$ on the sulfur vacancy ($V_S$). a) The hydrogen molecule approaches to the MoS$_2$ surface; b) The dissociation of $H_2$ is induced by the presence of a sulfur vacancy; c) Adsorption of H atoms in defective sites, forming H-Mo bonds.

On the other hand, DFT studies indicate that molecular hydrogen adsorption is more favorable in the MoS$_2$ surface if sulfur defective sites are present (**Figure 4**a), exhibiting a more effective energetic scenario than in a perfectly stoichiometric MoS$_2$ surface,[24,25] in accordance with other works that study the catalytic properties of MoS$_2$.[37,38] In this case, after $H_2$ dissociation (Figure 4b), H atoms bond to unsaturated molybdenum atoms,[25,26] as illustrated in Figure 4c. Considering the arguments presented above, and the finding that $V_S$ is generated in the mechanical exfoliation carried out for the preparation of MoS$_2$ monolayers,[40] we suggest that the electron donation from the hydrogen molecules probably originates from the dissociative adsorption of $H_2$ due the presence of $V_S$.

To investigate the role of $V_S$ in hydrogen adsorption, we cover the MoS$_2$ surface with aluminum oxide (Al$_2$O$_3$) layers using atomic layer deposition (ALD) and monitor the response of such hybrid MoS$_2$/Al$_2$O$_3$ FET to $H_2$. Different coverages of Al$_2$O$_3$ are directly deposited on the top of the MoS$_2$ device at 373 K, using TMA (Trimethyl Aluminum) and H$_2$O precursors. Previous works showed that the $Al_2O_3$ grows preferentially on terrace edges and localized defects, where the covalent lattice is discontinued breaking the surface periodicity.[41,42] Therefore, the $Al_2O_3$ clusters bind the intrinsic defects, including the $V_S$ in the $MoS_2$ monolayer. Based on the assumption that the sensor response depends on the dissociative adsorption of hydrogen in the $V_S$, we expect to see a lower sensitivity to $H_2$ molecules after the $Al_2O_3$ deposition. More details about the ALD steps to produce the $MoS_2/Al_2O_3$ FET and the AFM image before and after the $Al_2O_3$ grown are available in the Supplementary Information.

In **Figure 5**a we show the sensor response of a $MoS_2$ FET submitted to 5 growth pulses of $Al_2O_3$ and the bare $MoS_2$ device. This number of cycles was adopted to avoid the formation of an $Al_2O_3$ film on the $MoS_2$ device, solely promoting the passivation of $V_S$. The dotted curves represent the standard current, and the solid curves represent the current after 60 min of $H_2$ exposure at 473 K and $[H_2]$ = 20 %. According to these transfer curves, we observe that the current of $MoS_2/Al_2O_3$ FET (red curves), increases compared to the initial values for the bare $MoS_2$ FET (black curves). The FET mobility also increases from approximately 34 to 52 cm$^2$/V.s. This behavior is consistent with the dielectric screening effects from the $Al_2O_3$ overlayer observed in previous studies reported in the literature.[43]

We should point out that we continue to observe $H_2$ interaction after the $Al_2O_3$ growth. Most importantly, we notice a reduction in the charge transfer from hydrogen to the $MoS_2$ channel, indicating a decrease in the number of active sites of the $MoS_2$ surface for the $H_2$ interaction. The charge transfer for the bare $MoS_2$ FET is approximately $n$ ~7.4x10$^{11}$ cm$^{-2}$V$^{-1}$ while for the $MoS_2/Al_2O_3$ FET is $n$ ~ 3x10$^{10}$ cm$^{-2}$V$^{-1}$. We estimate the charge transfer by the $V_{TH}$ shift obtained by the extrapolation of the linear region between -17 V to -20 V, for the $I_{SD}$ vs. $V_G$ curves presented in Figure 5a. This behavior also occurs for devices in which a large amount of $Al_2O_3$ pulses is used.

To demonstrate the reduction of the $MoS_2$ reactivity to $H_2$ after the $Al_2O_3$ growth, we present in Figure 5b the current gain ($I_{H2}/I_{Ar}$) as a function of the threshold voltage difference ($V_G$-$V_{TH}$), this normalization again was adopted because $V_{TH}$ changes drastically after the oxide growth. The current gain is lower for the $MoS_2/Al_2O_3$ hybrid device represented by the red curve in Figure 5b than the bare $MoS_2$ device (black curve). Besides, $I_{H2}/I_{Ar}$ for $MoS_2/Al_2O_3$ FET is close to 1, indicating a small increase in current compared to its initial value without $H_2$ exposure. Additionally, the inset in Figure 5b shows the $S$ decrease after the $Al_2O_3$ growth. Such reductions in the current gain and sensor response demonstrate partial passivation of the reactive sites to hydrogen dissociative adsorption in the monolayer, evidencing that the $V_S$ play an important role in the $H_2$ interaction with the $MoS_2$. A similar decrease of the sensitivity of the $H_2$ detection system using edge-oriented vertically aligned $MoS_2$ flakes was observed after the passivation of the edges using ZnO films.[23]

Nonetheless, $MoS_2$ interaction with hydrogen continues to occur even after $Al_2O_3$ growth. Since $H_2$ is a small molecule and can reach both sides of the $MoS_2$ sheet (top and bottom side), one cannot disregard that hydrogen also interacts with the bottom side of the $MoS_2$ layer, producing the hydrogen response. This observation is also consistent with the high response and desorption times obtained for the $MoS_2$ FET sensors. The top side and bottom side of the monolayer are not equally accessible to $H_2$ molecules. Thus, two different mechanisms of interaction occur (see Fig S7 in SI). We believe that the fast increase (decrease, in case of the desorption process) of the current is explained by the reaction of $H_2$ with the top of the monolayer, while the longer reponse time for current saturation (or recovery) is explained by the diffusion of $H_2$ molecules in between $MoS_2$ and the substrate. Furthermore, the even higher $H_2$ desorption time in comparison with the $H_2$ adsorption time is related with the state of the "trapped" $H_2$ molecules in between the $MoS_2$ and the substrate having more difficult to be desorbed. Interestingly, the response and recovering time in the $MoS_2/Al_2O_3$ FET is longer than the bare $MoS_2$, as shown in the red curve of Figure 5c. This result is consistent with mostly the diffusion of hydrogen between the $MoS_2$ sheet and the $SiO_2$/Si substrate, and the attenuation of the fast adsorption mechanism at the top of the monolayer.

Finally, our work shows that the modulation of the conductivity of the $MoS_2$ monolayer channel dominates the $H_2$ sensing process. We propose that such modulation originates from a charge transfer process associated with the dissociative $H_2$ adsorption. Such adsorption is a thermally activated process (as shown in Figure 2a) that is facilitated by the presence of $V_S$ in the monolayer.[24,25] The effectiveness of $V_S$ to induce the charge transfer reaction implies that there is a limited number of sites for the reaction to occur, corroborating our results of charge-transfer saturation, as shown in Figure 1c. After aluminum oxide passivation of the top side of the monolayer, we yet detect hydrogen, but the sensor response decreases. This result is a clear indication that the $V_S$ plays an important role in the $H_2$ interaction with the $MoS_2$ and corroborates the idea that the gas molecule interacts with both sides of the $MoS_2$ sheet. Therefore, we believe that the sensor performance can be enhanced by the increase of $V_S$ on the $MoS_2$ surface. For instance, Donarelli *et al.* [44] proposed the creation of sulfur vacancies of MoS2 by thermal annealing in an ultra-high vacuum. An increase in the device mobility due to the $H_2$ presence is also observed that can be attributed to the screening of $MoS_2$ defective sites due to the hydrogen molecules adsorption. Besides, we believe that the existence of $V_S$ can contribute to the instability of the hydrogen adsorption, corroborated by the reversibility behavior of the reaction.

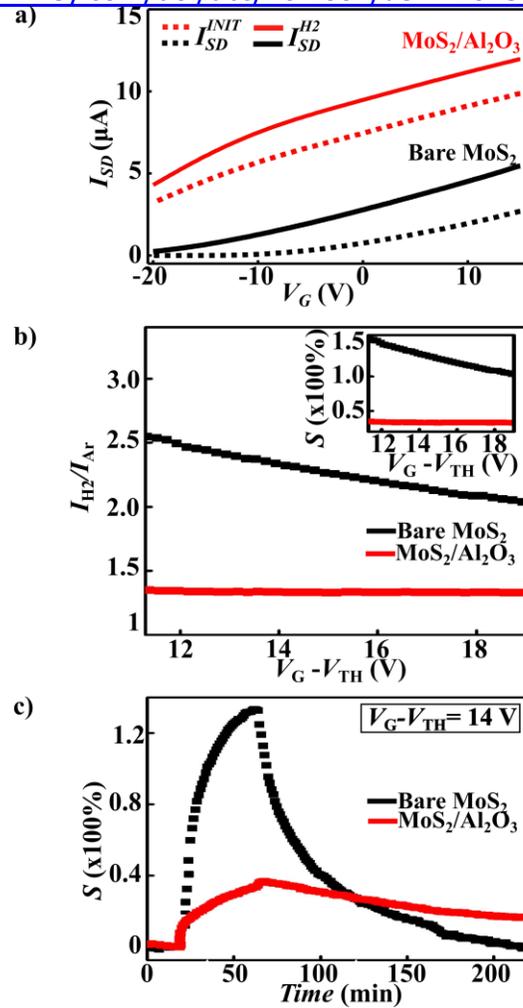

**Figure 5.** Comparison of sensing performance of $MoS_2$ monolayer and $MoS_2/Al_2O_3$ transistors are shown in a), b) and c), all studied for a fixed $[H_2]$ = 20 %, $V_{SD}$=1 V and 473 K. a) $I_{SD}$ vs. $V_G$ curves of the bare $MoS_2$ (black color), and $MoS_2$ with 5 cycles of $Al_2O_3$ growth (red color) before (dotted line) and after (solid lines) the $H_2$ exposure. b) Current gain ($I_{H2}/I_{Ar}$) as a function of the threshold voltage difference ($V_G$-$V_{TH}$) for bare $MoS_2$ and $MoS_2/Al_2O_3$ FETs. Inset: Sensor response as a function of the threshold voltage difference. c) Sensor response as a function of time during the $H_2$ exposure, in fixed threshold voltage difference ($V_G$-$V_{TH}$) = 14 V.

## 3. Conclusion

We investigated the $H_2$ interaction with monolayer $MoS_2$ devices. Both the adsorption and desorption gas processes are strongly dependent on the temperature, and the sensing is independent of the choice of the substrate or metallic contacts. Our experimental data suggest that the charge transfer process from $H_2$ to the $MoS_2$ channel dominates the sensing process. We propose that the charge transfer originates from the dissociative $H_2$ adsorption, which is facilitated by the presence of $V_S$ in the $MoS_2$. We also discussed evidence that the molecular hydrogen can interact on both sides of the $MoS_2$ sheet (top side and bottom side) based on our measurements of $MoS_2/Al_2O_3$ heterojunctions. Finally, we demonstrated that the $MoS_2$ device is a promising candidate for the development of molecular hydrogen sensors. The $MoS_2$ sensor responds to a wide range of hydrogen concentrations, operates at relatively low temperatures and do not require the presence of catalytic metals dopants (Pd, Pt), and can be fully recovered. Our findings also provide insights for future applications in catalytic processes making use of monolayer $MoS_2$ FETs.

## 4. Experimental Section

*Device Fabrication:* To investigate the molecular hydrogen interaction with $MoS_2$ FET, we produce several devices supported on two different substrates: Si covered with 285 nm of $SiO_2$ ($SiO_2/Si$) and hexagonal boron nitride (h-BN). To obtain the monolayer $MoS_2$, we used mechanical exfoliation with scotch tape technique.[45] To prepare the $MoS_2$ FET supported on h-BN, firstly, the mechanically exfoliated h-BN flakes are directly transferred from the tape to the $SiO_2/Si$ substrate. After that, we use the dry viscoelastic stamping technique to transfer the $MoS_2$ in the top of h-BN/$SiO_2$/Si flakes.[46] The metallic contacts of both device types are fabricated employing standard electron beam lithography and thermal metal deposition of Au (50nm). Finally, we use a second lithography step and $SF_6$ plasma etching to define the $MoS_2$ device geometry.

*In situ Electrical Measurements:* After fabrication, the $MoS_2$ FET is fixed on a chip holder and transferred into a chamber connected to an electrical measurement system. We use a heater to control the temperature inside the chamber in the range of 300 K–473 K. The $H_2$ flow is determined by a mass flow controller and dilution with ultra-high pure argon (Ar) is used to obtain different $H_2$ concentrations. More details of the gas sensing system can be found in our previous work.[7] Before carrying out electrical measurements under hydrogen exposure, we perform an annealing procedure in all devices presented in this work in Ar atmosphere at 473 K for 12 hours. This thermal treatment is known to promote the removal of contaminating gases and humidity.[7,31,32] We use the external DC source of a lock-in amplifier (SR830) to provide a source-drain voltage ($V_{SD}$) and a Keithley K2400 source to provide a DC gate voltage ($V_G$). The current between the source and drain ($I_{SD}$) is collected by a pre-amplifier and measured by a Keithley 2000 Digital Multimeter.

**Supporting Information**

Supporting Information is available from the Wiley Online Library or from the author.

**Acknowledgments**:

This work was supported by CAPES, Fapemig (Rede 2D), CNPq and INCT/Nanomaterials de Carbono. The authors are thankful Lab Nano at UFMG for allowing the use of atomic force microscopy and CT Nano at UFMG for the Raman measurements. We acknowledge S. L. M. Ramos for the Raman measurements and E. S. N. Gomes for helping with the Raman analyses. We also acknowledge M. S. C, Mazzoni for helping with theoretical discussions about phenomenology. K.W. and T.T. acknowledge support from the Elemental Strategy Initiative conducted by the MEXT, Japan and JSPS KAKENHI Grant Numbers JP15K21722.

This is the authors' version (pre peer-review) of the manuscript: N R. Rezende et al, Advanced Electronic Materials  https://doi.org/10.1002/aelm.201800591 , That has been published in its final form:    https://onlinelibrary.wiley.com/doi/abs/10.1002/aelm.201800591References

Supporting Information

**Probing the Electronic Properties of Monolayer MoS$_2$ via Interaction with Molecular Hydrogen**

*Natália P. Rezende, Alisson R. Cadore, Andreij C. Gadelha, Cíntia L. Pereira, Vinicius Ornelas, Kenji Watanabe, Takashi Taniguchi, André S. Ferlauto, Ângelo Malachias, Leonardo C. Campos, and Rodrigo G. Lacerda\**

1. **Evaluation of response time of MoS$_2$ FETs at 473 K for distinct *[H$_2$]***

Figure S1 shows the response time of a MoS$_2$ FET subjected to distinct H$_2$ concentrations at 473 K. The saturation current was estimated by the exponential fit of the $I_{SD}$ x *Time* curves shown in Figure 1e of the main text. The response time of a sensor decreases with the increase in *[H$_2$]*. This phenomenon may be related to the $T_{RES}$ definition, since an increase in the number of hydrogen molecules in the MoS$_2$ surface leads to a faster current saturation.

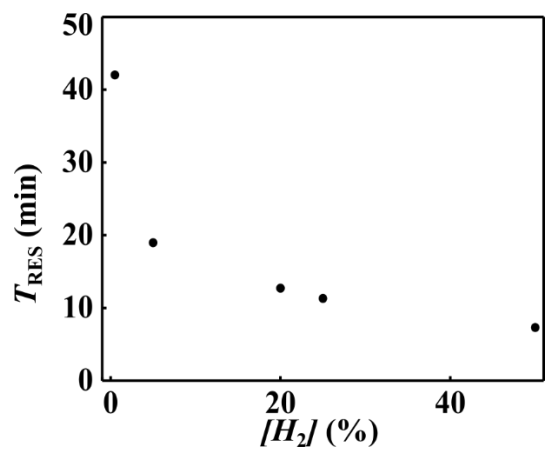

**Figure S1:** Response time of MoS$_2$ FETs for different H$_2$ concentrations of *[H$_2$]* at 473 K.

We found that under the same exposure conditions (temperature and $H_2$ concentrations) the response time of all produced devices is similar. We also found that the response time remains mostly unchanged for repeated measurements using the same device.

## 2. Reversibility and reproducibility of the $I_{SD}$ x *Time* curves for $H_2$ exposure at 473 K

To investigate the reversibility and reproducibility of the $H_2$ reaction with the monolayer, we carried out long $I_{SD}$ vs. *Time* measurements of the $MoS_2$ FET under 160 minutes exposure to 20 % $H_2$ at $V_G$= -20 V and temperature of 473 K, as shown in Figure S2. The red curve measurement was obtained right after the black curve, exhibiting only small changes in comparison with the first curve, probably due to experimental variations. These results indicate a good reproducibility of the device behavior. We observe a complete recovery in these conditions after 16 h in pure Ar.

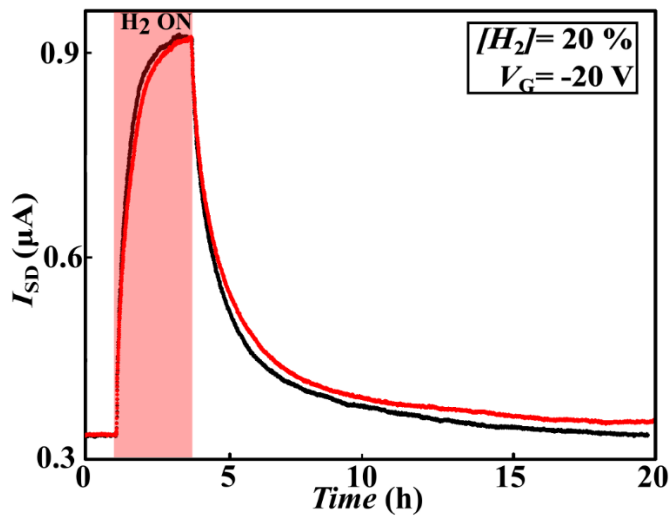

**Figure S2:** Source-drain current ($I_{SD}$) as a function of time for 160 min under 20 % $H_2$ exposure for $MoS_2$ FET at $V_G$= -20 V and 473 K. The red curve measurement was obtained immediately after the black curve measurement.

3. **Raman study of structural changes in the MoS$_2$ FET due to the H$_2$ exposure**

We searched for possible structural changes in the MoS$_2$ due to the H$_2$ exposure using Raman spectroscopy. The Raman spectra and maps were recorded using a WITec Alpha 300 system operating with a 532 nm excitation source at 0.5 mW power. To investigate the variations in the Raman spectra, we compare the spectra before and after the exposure of the MoS$_2$ to *[H$_2$]*= 65% at 473 K, during 6 h, as shown in Figure S3. The Raman spectra before the H$_2$ exposure are represented by blue curves in Figure S3a. The energy difference (Δ) between the $E^1_{2g}$ and $A_{1g}$ modes present a fixed value of Δ= 19 cm$^{-1}$, which is an accurate indicator of layer thickness for monolayer MoS$_2$.[1] In the Figure S3b, we show a comparison of the spatially resolved Raman spectroscopy of $E^1_{2g}$ (top Figure) and $A_{1g}$ (bottom Figure) modes before and after H$_2$ exposure. Both Raman spectra and maps were calibrated using the silicon peak after the H$_2$ exposure to reproduce the measurements. We do not observe relevant shifts in our experiments according to the precision of the measurement setup, of approximately 1 cm$^{-1}$.

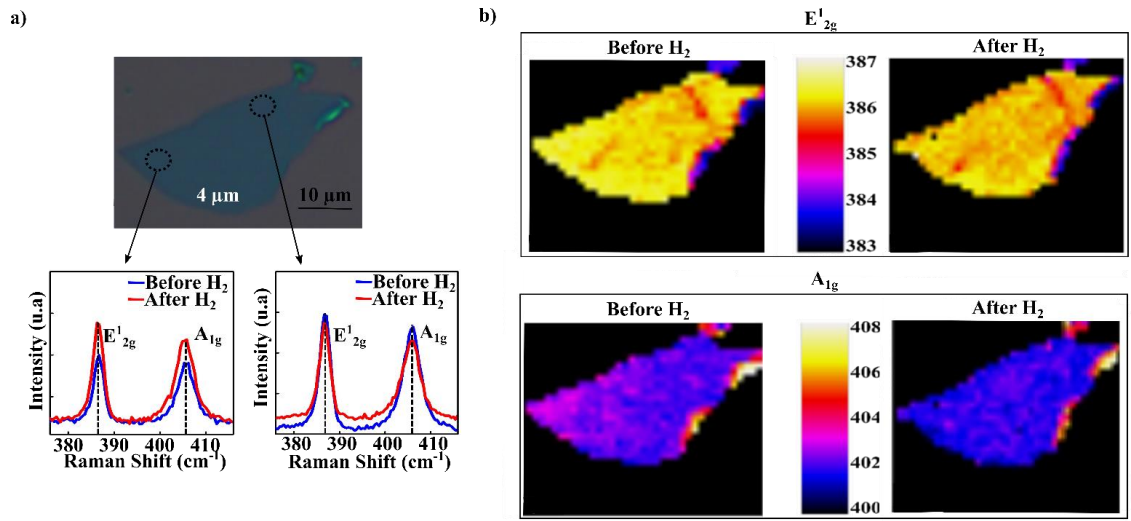

**Figure S3:** a) Raman spectra of the MoS$_2$ before and after the exposure to *[H$_2$]* = 65% at 473 K, during 6 h; b) Spatially resolved Raman spectroscopy of $E^1_{2g}$ and $A_{1g}$ modes before and after H$_2$ exposure.

## 4. Four probe measurement configuration

Figure S4a and S4b show a sketch of the circuits for two and four probe configuration, respectively, considering the resistances that influence the measurements. As depicted in Figure S4a, when we carry out electric measurements using two terminals, the contact resistances are included in the total resistance of the device, since each contact is used to both apply current and measure the voltage. To avoid the effect of the contact resistance on the measurements, we performed measurements with four terminals, as shown in Fig. S4b. To carry out the four probe measurements, we used a Keithley K2400 source to applied a fixed DC current ($I$) of $1 \times 10^{-6}$ A between the source ($Sr$) and drain ($D$) terminals. The voltage ($V$) between the internal terminals $C_1$ and $C_2$ (Figure S4b) was measured by a Keithley 2000 Digital Millimeter. A Keithley K2400 source was used to provide a DC gate voltage ($V_G$).

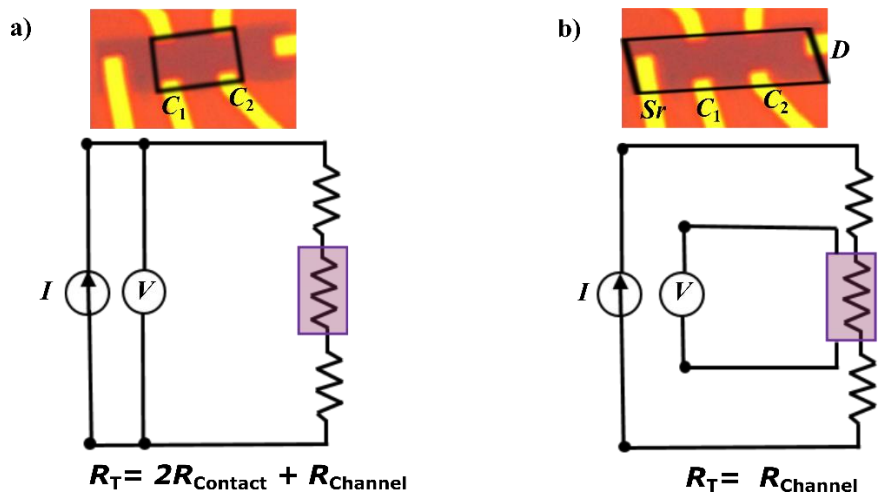

**Figure S4:** Schematic representation of the circuits for two (a) and four probe (b) configurations, considering the resistances that influence the electric measurements.

## 5. MoS₂ FET supported on h-BN under H₂ exposure

Similarly to the measurements that were carried out for the $MoS_2$ FET supported on $SiO_2$, we investigated the response of the $MoS_2$ FET supported on h-BN (Figure S5a) to the $[H_2] = 20\%$ at 473 K (Figure S5b). Also, we measured a wider range of concentrations, spanning from 0.1 to 90% of $H_2$, as shown in Figure S5c. We must emphasize here that the devices have the same response behavior as a function of temperature and $H_2$ concentration was observed for both substrates.

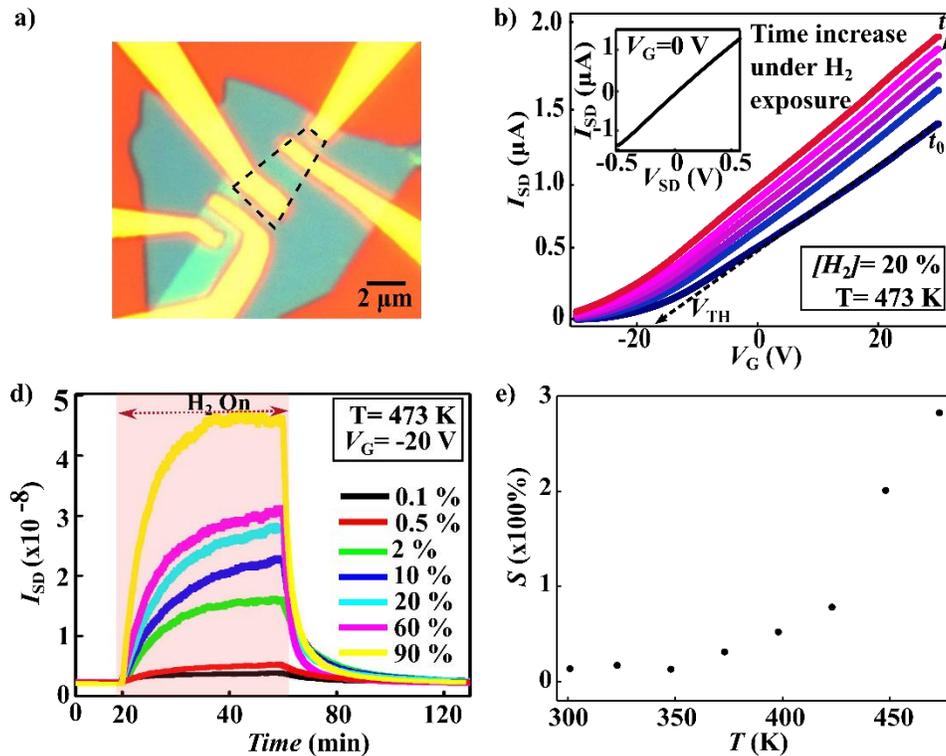

**S5:** Study of $H_2$ interaction with a $MoS_2$ FET supported on h-BN at 473 K and $V_{SD}$ = 0.1 V. a) Optical image of a typical $MoS_2$/BN device; b) $I_{SD}$ x $V_G$ curves of the $MoS_2$/BN transistor under 20% of $H_2$ exposure. Inset: $I_{SD}$ x $V_{SD}$ curve for $V_G$ = 0 V, before $H_2$ exposure; c) Current change versus time in different hydrogen concentrations in fixed $V_G$= -20 V; d) Sensor response of $MoS_2$ monolayer as a function of temperature: from 300 K up to 473 K. The data were taken at a fixed gate voltage $V_G$= -20 V.

## 6. Atomic layer deposition (ALD) of $Al_2O_3$ on $MoS_2$ devices

### 6.1- ALD technique and $MoS_2/Al_2O_3$ morphology

The atomic layer deposition (ALD) of $Al_2O_3$ consists of two processes. The first ALD treatment is carried out with 50 trimethylaluminum (TMA) pulses (the Al precursor), which react with available terminations (vacancies, edges and point defects) in $MoS_2$. Next, alternated cycles of TMA (0.015s) and $H_2O$ (0.015s) were used (with 30 s purge time after each pulse) to obtain the $MoS_2/Al_2O_3$ FETs. The calibrated thickness produced for a single TMA/$H_2O$ cycle is ~0.09 nm.[2,3] During the growth-process of $Al_2O_3$ films, the substrate temperature was fixed at 423 K.

In Figure S6 we show the AFM image before and after 100 cycles of $Al_2O_3$ grown on a monolayer $MoS_2$ to evidence the morphology of the oxide film. The $Al_2O_3$ film is highly non-uniform due to the preferential nucleation at edges and defects. The $Al_2O_3$ clusters grow in the shape of nanospheres, which can be explained by the low surface energy of monolayer $MoS_2$ due to the presence of few effective dangling bonds.[3,4]

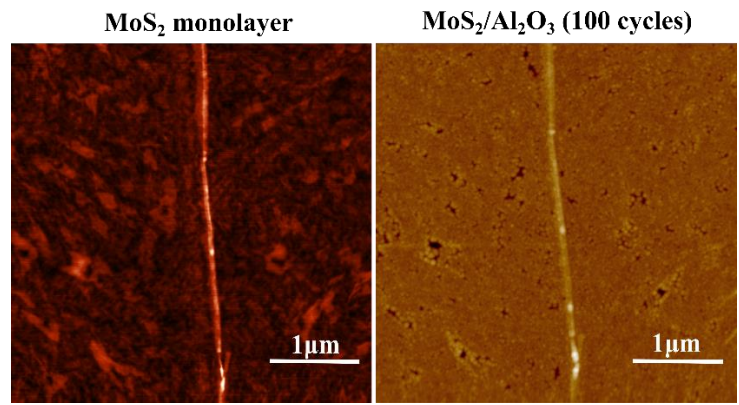

**Figure S6:** AFM image of the monolayer $MoS_2$ without $Al_2O_3$ and the $MoS_2/Al_2O_3$ hybrid structure

**6.2- Mechanisms of adsorption and desorption of the $H_2$ in the $MoS_2$ FET**

We believe that the fast increase in the current (region 1, in the **Figure S7**a) is explained by the reaction of $H_2$ with the top of the monolayer, while the longer time for the current saturation (region 2, in the Figure S7a) is explained by the diffusion of $H_2$ in between $MoS_2$ and the substrate. After the passivation of S vacancies in the $MoS_2$ with $Al_2O_3$ growth, both the response and recovery time increased due to the attenuation of fast adsorption and desorption mechanism at the top of the monolayer, as shown in Figure S7b.

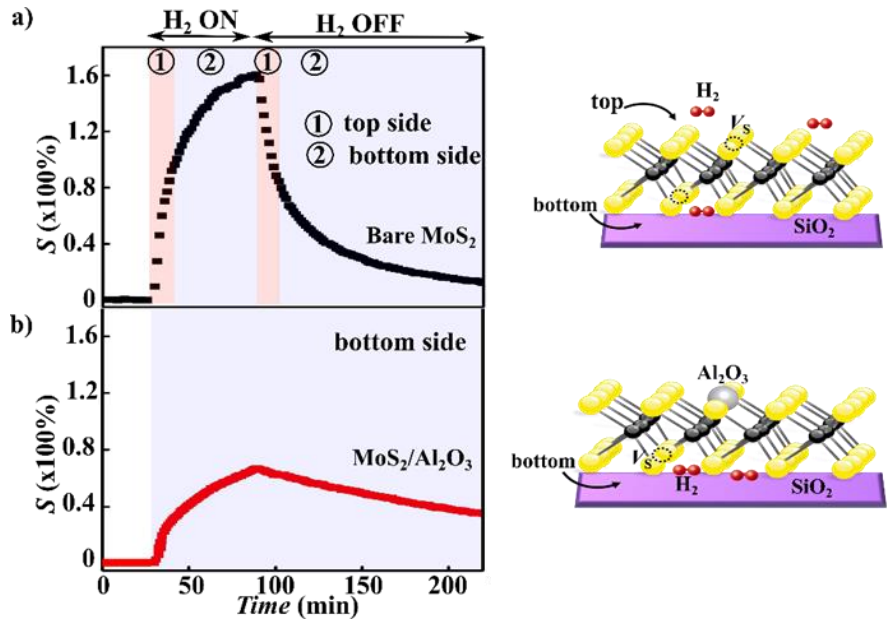

**Figure S7:** Sensor response as a function of time during the $H_2$ exposure, in fixed threshold voltage difference $(V_G\text{-}V_{TH}) = 14$ V for the bare $MoS_2$ FET (a) and $MoS_2$ with 5 cycles of $Al_2O_3$ growth (b). The region 1 and 2 represent the interaction of $H_2$ with the top side and bottom side respectively. After the passivation of S vacancies in the $MoS_2$ with $Al_2O_3$ growth occurs the attenuation of fast adsorption mechanism at the top of the monolayer.